# Black Holes at high and low metallicity

**Jorick S. Vink, Gautham N. Sabhahit, and Ethan R.J. Winch**

Armagh Observatory and Planetarium

**Abstract.** At the end of their lives the most massive stars collapse into black holes (BHs). The detection of an 85 $M_\odot$ BH from GW 190521 appeared to challenge the upper-mass limit imposed by pair-instability (PI). Using systematic MESA calculations with new mass-loss implementations, we show that 100 $M_\odot$ stars at metallicities below 0.1 $Z_\odot$ can evolve into blue supergiant progenitors with cores small enough to avoid PI, yet with limited envelope loss, yielding remnants within the second mass gap. The key ingredients involve (i) a proper consideration of internal mixing and (ii) physically motivated stellar winds. Our modelling provides a robust pathway that roughly doubles the maximum BH mass permitted by PI theory and establish a physically-consistent framework to explore the upper BH mass limit versus metallicity.For rapid rotation ($\geq$50% of critical), the upper BH mass comes down to $\simeq$35 $M_\odot$, matching the LIGO/Virgo BH mass pile-up.

---

## 1. Introduction

Stars form with a range of masses ($M$), metallicities ($Z$), rotation rates ($\Omega$), and mixing parameters (e.g. described by the overshooting parameter $\alpha_{\rm ov}$). The most massive stars end their lives as BHs but their final mass depends sensitively on stellar wind mass loss, which in turn is a function of $Z$.

In this short contribution, we summarise recent work by the Armagh Massive Star Evolution Group† on the final black hole (BH) masses predicted at both high (Galactic) and low metallicities ($Z$). This work is largely motivated by the detection of heavy BHs with LIGO/Virgo, but its implications extend far beyond gravitational-wave astrophysics. In particular, determining whether very massive stars (VMS) undergo pair-instability (PI) supernovae – releasing large quantities of metals – or instead collapse directly into BHs without enriching their surroundings has major consequences for models of galactic chemical evolution (GCE).

## 2. The Maximum Black Hole mass at Galactic $Z$

In high Galactic $Z$ environments mass-loss rates are large. While there are still many uncertainties in stellar wind mass-loss rates, the situation around $\sim$80-100 $M_\odot$ is surprisingly good, as we can avail of the transition mass-loss rate of Vink & Gräfener (2012). At this point, winds are expected to switch from being optically thin to optically thick – just when the winds become sufficiently dense for multiple scattering to take over. This results in an upturn or kink of the mass-loss rate versus luminosity or Eddington factor $\Gamma$ (proportional to $L/M$) as shown by Vink et al. (2011) and Sabhahit et al. (2025c).

While general empirical mass-loss rate determinations are affected by uncertainties due to wind clumping, the transition mass-loss rate is largely independent of such details. It is determined primarily by the stellar luminosity, which is far better constrained than mass-loss rates that can be uncertain by up to an order of magnitude.

Applying the Vink et al. (2011) kink in the stellar evolution code MESA (Paxton et al. 2013), Sabhahit et al. (2022) found that VMS essentially evaporate to much smaller masses than their initial zero-age main sequence (ZAMS) values. For example, starting from a $150\,M_\odot$ VMS on the ZAMS, the models evolve almost vertically downward in the Hertzsprung–Russell diagram, decreasing in both mass and luminosity until less than 10% of the initial mass remains.

† Space limitations prevent us from providing additional context and references to related studies, for which we apologise.

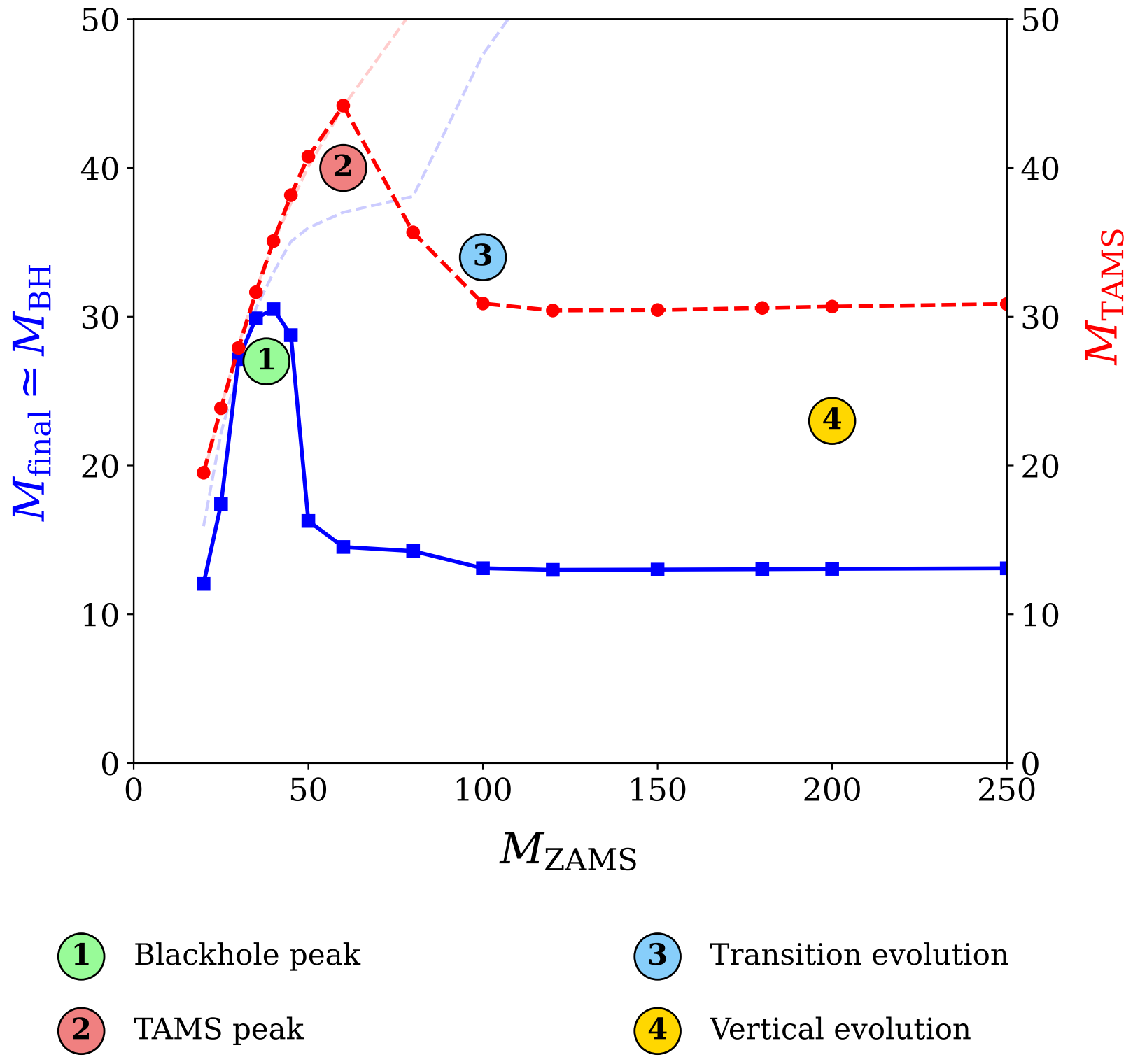

**Figure 1.** Final black-hole (BH) mass as a function of ZAMS mass (blue). The red dashed line denotes the TAMS mass. Adapted from Vink et al. (2024).

As a result of this implementation, Vink et al. (2024) assessed the maximum BH mass at Galactic (solar) metallicity, finding it to be on the order of $30\,M_\odot$ (see Fig. 1). Interestingly, the peak of this maximum BH mass does not occur at the highest stellar masses, but rather at comparatively low initial masses, below $50\,M_\odot$ (see the blue line). Moreover, the shape of the BH mass distribution closely resembles that of the terminal-age main sequence (TAMS) mass distribution (red dashed line), but transformed downward from slightly higher initial masses.

In other words, the shape is determined not by what happens during the evolved phases of a (very) massive star's life, but by what occurs at the very onset of core hydrogen (H) burning.

## 3. Maximum BH mass at low metallicity

The other crucial parameter determining a star's fate is the degree of internal mixing, particularly during core H burning. Until around 2010, most stellar evolution models adopted relatively small amounts of core overshoot ($\alpha_{\rm ov} \leq 0.1$). However, to address the unexpectedly large number of B supergiants revealed by the VLT-FLAMES data, we explored higher overshoot values in the range $\alpha_{\rm ov} = 0.3$–$0.5$ (Vink et al. 2010). For this reason, we now allow for a distribution of overshooting parameters for each and every stellar mass (e.g. Higgins & Vink 2019).

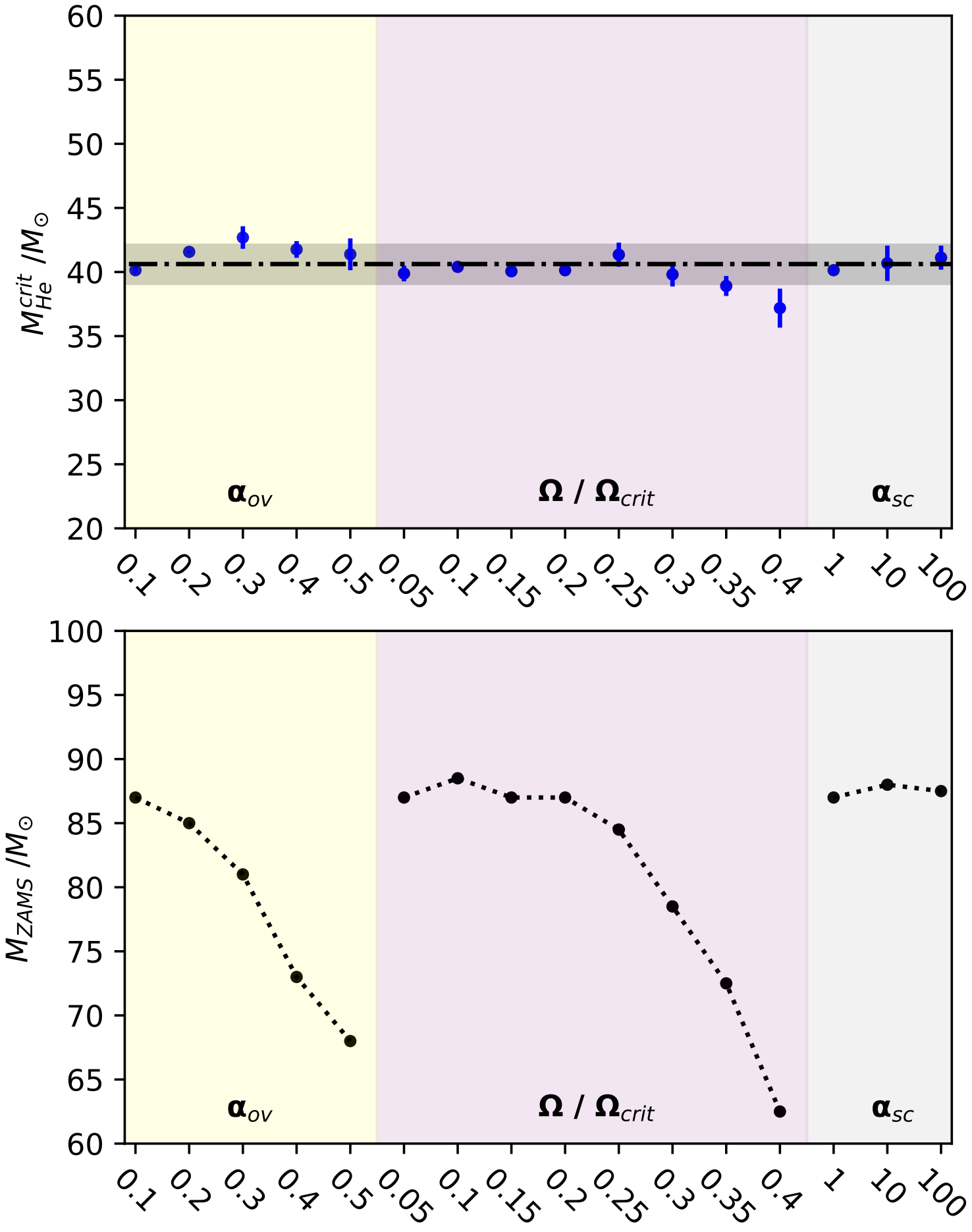

**Figure 2.** Top panel: Critical He-core mass as a function of three independent internal mixing parameters ($\alpha_{\rm ov}$, $\Omega$, and the semi-convection parameter $\alpha_{\rm sc}$). Bottom panel: Initial ZAMS masses that these models correspond to. Overshoot and rotation have a significant impact on the initial mass, while semi-convection is less influential. Adapted from Winch et al. (2024).

So when the spectacular event GW 190521 occurred, placing the primary BH firmly within the PI gap, we were ready not only to ask but also to address whether stellar evolution models could indeed produce such heavy $\sim$85 $M_\odot$ black holes.

Indeed, for stars in the initial mass range 85–100 $M_\odot$, the stars were massive enough—but not so massive as to develop oversized convective cores that would exceed the critical CO-core mass of 36 $M_\odot$ (or the corresponding He-core mass of 40 $M_\odot$). As long as internal mixing was kept to a minimum, stars at low $Z$ could retain large envelope masses (owing to weak wind mass loss), evolve as blue supergiant (BSG) progenitors, and ultimately produce 85–90 $M_\odot$ black holes (Vink et al. 2021). For higher degrees of overshooting and/or rotation, this upper limit would naturally come down.

To perform a more systematic analysis, Winch et al. (2024) extended the study of Vink et al. (2021) by varying a wider range of parameters. An overview of their results is shown in Fig. 2. Winch et al. (2024) first carried out a series of critical core mass experiments, demonstrating that the critical He- and CO-core masses are largely independent of internal mixing parameters such as $\alpha_{\rm ov}$, $\alpha_{\rm sc}$, or $\Omega/\Omega_{\rm crit}$ (see the top panel of Fig. 2).

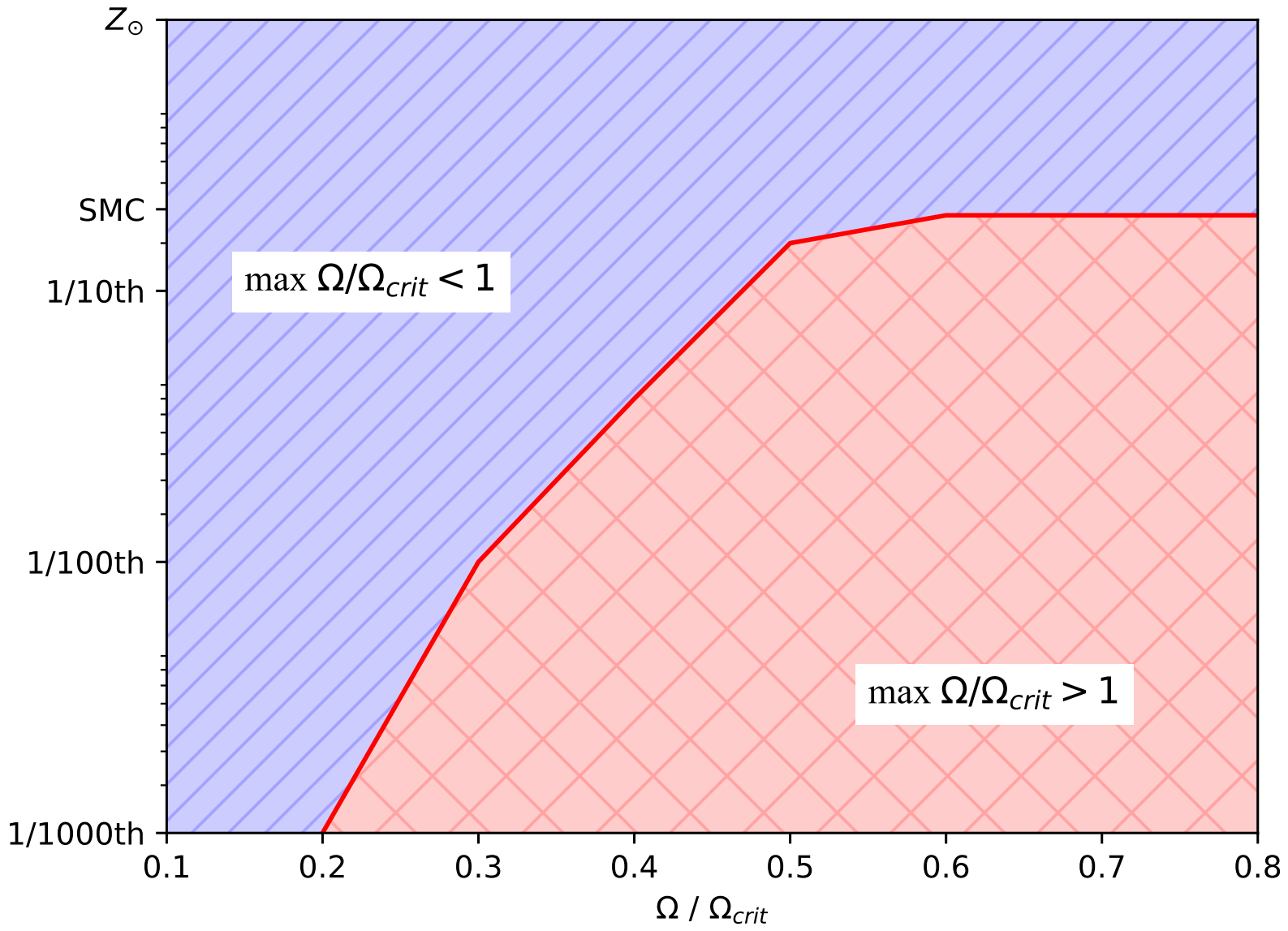

**Figure 3.** At SMC and higher metallicities, critical rotation (break-up) is not reached. At lower $Z$, break-up is attained progressively at lower $\Omega$. Adapted from Winch et al. (2025).

As a second step, a sample of models 100 times larger than that of Vink et al. (2021) was computed to determine the corresponding initial stellar and final BH masses (see the bottom panel of Fig. 2). We subsequently performed population synthesis based on these models and found good agreement with the observed shape of the LIGO BH mass function, including its natural cut-off around 85–90 $M_\odot$ in O3.

## 4. The Role of rapid rotation

While it is possible to produce such massive BHs in slowly rotating models, rapidly rotating stars may behave quite differently. In Winch et al. (2025), we therefore investigated the effects of rapid rotation by implementing the GENEC rotational mass-loss prescription (Maeder & Meynet 2000) into MESA. We also accounted for whether the stars would reach critical break-up rotation (see Fig. 3) and included an additional mass-loss prescription for that regime.

As shown by the parameter extension to higher $\Omega$ in Fig. 4, the final BH masses (green line) converge to a similar value. After performing population synthesis on this extended model grid, we concluded in Winch et al. (2025) that our rapidly rotating models naturally reproduce the 35 $M_\odot$ mass pile-up feature observed in the LIGO/Virgo BH mass distribution. Although these stars also evolve chemically homogeneously and may appear as stripped stars, the mass feature itself arises from the mass loss, rather than from chemically homogeneous evolution (CHE) per se. The two phenomena just happen to occur simultaneously.

## 5. Summary

We performed stellar evolution calculations of (very) massive stars across a wide metallicity ($Z$) range using MESA, exploring various internal mixing parameters, to assess the maximum BH mass attainable at both high and low $Z$. Our main conclusions are as follows:

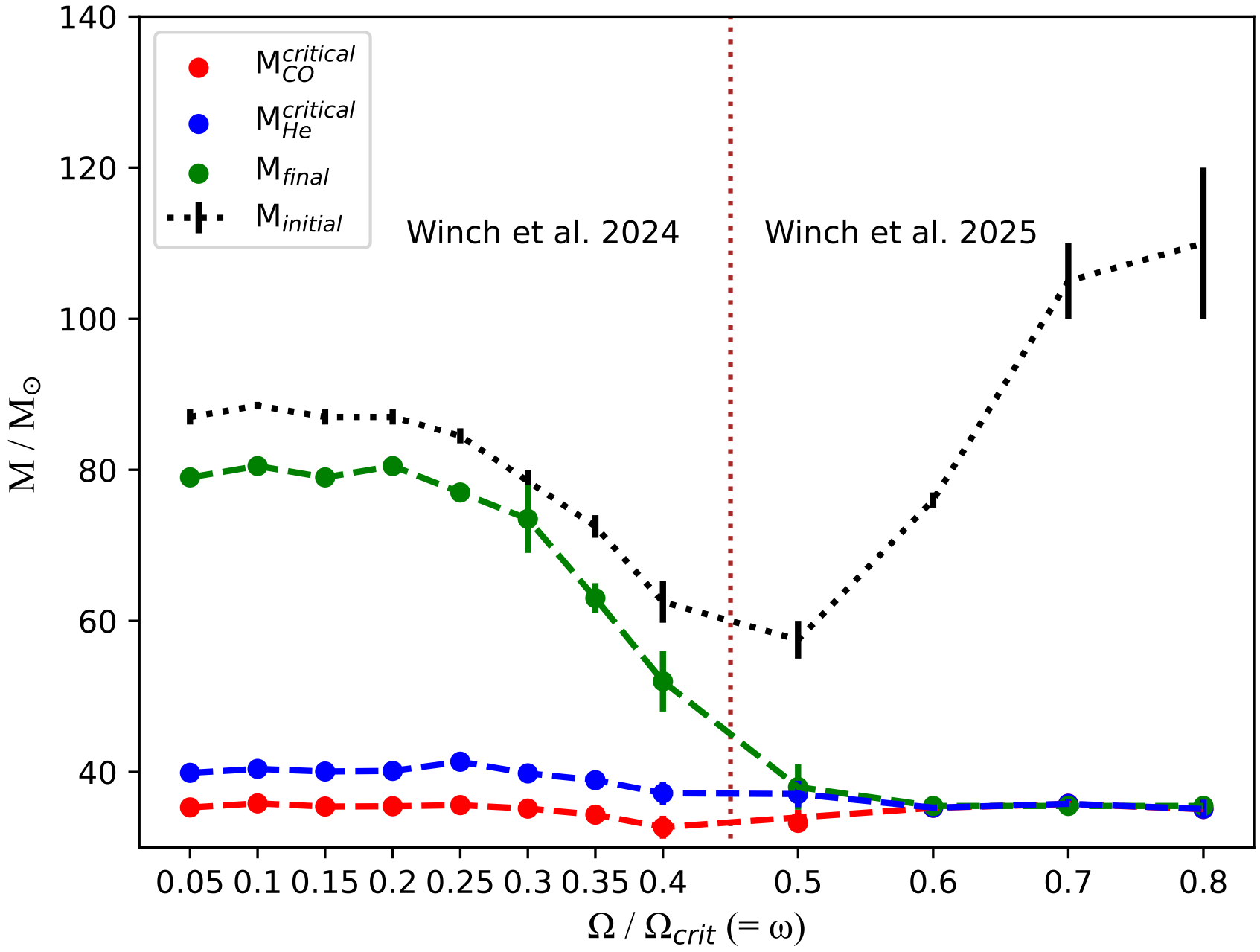

**Figure 4.** Comparison between the rapidly rotating models of Winch et al. (2025) and the slowly rotating models of Winch et al. (2024).

• At solar $Z$, the maximum BH mass is of order $30\,M_\odot$. If significantly heavier BHs are discovered in the Milky Way, they likely originated from an epoch when the Galaxy had a lower $Z$, as was probably the case for Gaia BH-3.

• For stars below the pair-instability threshold, the maximum black hole mass almost doubled from the traditionally assumed $\sim 50\,M_\odot$ to approximately $90\,M_\odot$.

• Rapid rotation lowers these values to around $35\,M_\odot$, corresponding to the distinct bump seen in the LIGO/Virgo black hole merger mass distribution. In other words, the $35\,M_\odot$ pile-up feature observed by LIGO may arise from objects rotating at more than 50% of their critical rate.